\documentstyle [twocolumn,epsf]{mn}
\newcommand{\mincir}{\raise
-3.truept\hbox{\rlap{\hbox{$\sim$}}\raise4.truept\hbox{$<$}\ }}
\newcommand{\magcir}{\raise
-3.truept\hbox{\rlap{\hbox{$\sim$}}\raise4.truept\hbox{$>$}\ }}
\newcommand{\minmag}{\raise
-3.truept\hbox{\rlap{\hbox{$<$}}\raise5.truept\hbox{$<$}\ }}
\newcommand{\be}{\begin{equation}}
\newcommand{\ee}{\end{equation}}
\newcommand{\CC}{\Lambda}
\newcommand{\rLo}{\rho_{\CC}^0}
\newcommand{\rL}{\rho_{\CC}}
\newcommand{\rmr}{\rho_m}
\newcommand{\pmr}{p_m}
\newcommand{\wm}{\omega_m}
\newcommand{\ba}{\begin{eqnarray}}
\newcommand{\ea}{\end{eqnarray}}
\newcommand{\brr}{\begin{array}}
\newcommand{\err}{\end{array}}
\newcommand{\bc}{\begin{center}}
\newcommand{\ec}{\end{center}}

\renewcommand{\baselinestretch}{1.0}

\title[Expansion History with Decaying Vacuum]
{Expansion History with Decaying Vacuum: A Complete Cosmological Scenario}
\author[J. A. S. Lima, S. Basilakos \& Joan Sol\`a]{J. A. S. Lima$^1$,
S. Basilakos$^2$, Joan Sol\`a$^{3}$\\
\vspace{0.1cm} $^1$ Departamento de Astronomia, Universidade de S\~ao
Paulo, Rua
do Mat\~ao 1226, 05508-900, S\~ao Paulo, Brazil\\
$^2$ Academy of Athens, Research Center for Astronomy \& Applied
  Mathematics, Soranou Efessiou 4, 11-527, Athens, Greece\\
$^3$ High Energy Physics Group, Dept. ECM, and Institut de
Ci\`encies del Cosmos (ICC), Univ. de Barcelona, Av. Diagonal 647\\
 E-08028 Barcelona, Catalonia, Spain
}

\begin{document}

\maketitle

\begin{abstract}
We propose a novel cosmological scenario with the space-time emerging from
a pure initial de Sitter stage and subsequently evolving into the
radiation, matter and dark energy dominated epochs, thereby avoiding the
initial singularity and providing a complete description of the expansion
history and a natural solution to the horizon problem. The model is based
on a dynamical vacuum energy density which evolves as a power series of
the Hubble rate. The transit from the inflation into the standard
radiation epoch is universal, giving a clue for a successful description
of the graceful exit. Since the resulting late time cosmic history is very
close to the concordance $\Lambda$CDM model, the new unified framework
embodies a more complete past cosmic evolution than the standard
cosmology.

\end{abstract}

\vspace{1.0cm}

\section{Introduction}
In the current view of the cosmological history it is believed that matter
and space-time emerged from a singularity and evolved through four
different eras: early inflation, radiation, dark matter (DM) and dark
energy (DE) dominated eras. During the radiation and DM dominated stages,
the expansion of the Universe slows down while in the inflationary and DE
eras it speeds up. So far there is no clear cut connection between the
inflationary period and the normal Friedmann expansion. Moreover the
current DE phase is also a mystery.

Over the past decade, studies of the available high quality cosmological
data (supernovae type Ia, CMB, galaxy clustering, etc.) have converged
towards a cosmic expansion history that involves a spatially flat geometry
and a recent accelerating period of the Universe (cf. Spergel et al. 2007;
Amanullah et al. 2010; Komatsu et al. 2011 and references therein). This
faster expansion phase has been attributed to the DE component with
negative pressure. The simplest type of DE corresponds to the cosmological
constant (hereafter CC) (Weinberg 1989; Peebles \& Ratra 2003; Padmanabhan
2003; Lima 2004). The so-called concordance model (or $\Lambda$CDM model
(Peebles 1993), which contains cold DM to explain clustering, flat spatial
geometry and a CC, $\Lambda$, fits accurately the current observational
data and thus it is an excellent candidate to be the model that describes
the observed Universe. However, the $\Lambda$CDM suffers from, among
others, two fundamental problems: (a) The ``old'' cosmological constant
problem (or {\it fine tuning problem}) i.e., the fact that the observed
value of the vacuum energy density is many orders of magnitude below the
value suggested in quantum field theory (QFT) (Weinberg 1989; Peebles \&
Ratra 2003; Padmanabhan 2003; Lima 2004), and (b) {\it the coincidence
problem} (see Steinhardt 1997) i.e., the fact that the (decreasing) matter
energy density and the (constant) vacuum energy density happen to be of
the same order just prior to the present epoch.

One of the main attempts to solve or at least to alleviate such
theoretical problems is based on the idea that the   vacuum energy density
is a time-dependent quantity, i.e. $\CC=\CC(t)$. There are a number of
interesting $\Lambda$-variable models in the old literature (Ozer \& Taha
1986, 1987; Bertolami 1986; Freese et al. 1987; Peebles \& Ratra 1988;
Carvalho, Lima \& Waga 1992; Lima \& Maia 1994; Lima 1996; Lima \& Trodden
1996; Overduin \& Cooperstock 1998) and even more recently (Shapiro \&
Sol\`a 2000; Shapiro \& Sol\`a 2002; Alcaniz \& Maia 2003; Opher \&
Pellison 2004; Bauer 2005; Carneiro \& Lima 2005; Alcaniz \& Lima 2005;
Barrow \& Clifton 2006; Montenegro \& Carneiro 2007; Shapiro \& Sol\`a
2009; Sol\`a \& \v{S}tefan\v{c}i\'{c} 2005, 2006; Basilakos 2009; Sol\`a
2011 and references therein). The functional form of $\CC(t)$ in most of
them has usually been proposed on phenomenological grounds, as it occurs
with the vast majority of DE models (Carvalho et al. 2006).  Apart the
paper by Lima \& Maia (1994) and Lima \& Trodden (1996) their basic
motivation was to understand the present day smallness of the vacuum
energy density, that is, with no attempt to find a clear correlation
between  the values of $\Lambda$ during inflation and at recent times.
This occurs because the functional form of $\Lambda(t)$ is chosen just by
extrapolating backwards in time the present available cosmological data,
including a divergent upper  bound  for $\Lambda$. However, since
$\Lambda_0$  is  most likely a remnant from inflation, a more realistic
phenomenological decaying  law should describe all cosmic history,
embodying the primordial inflation itself driven by a finite $\Lambda$.
Naturally,  a more fundamental approach for the decaying vacuum models
based on quantum field theory (QFT) or at least partially justified from
first principles should be desirable in physical grounds. A proposal
along these lines in which the presently observed $\Lambda$ appears as a
quantum relic from inflation has been discussed in the literature (Sol\`a
2008). A related work but in the context of a five dimensional
Noncompact Kaluza-Klein gravity model can be found in Anabitarte,
Aguilar \& Bellini (2006).

In the present paper we propose a new complete cosmological history based
on decaying vacuum models.  Our basic aim is to understand the possible
relationship between the late time magnitude of the $\Lambda$-term and its
upper bound during the inflationary epoch. As it will be seen, some
constraints from  QFT in curved spacetime are naturally incorporated in
our scenario, specially those coming from recent developments inspired in
the renormalization group approach (Shapiro \& Sol\`a 2002; Sol\`a 2008;
Shapiro \& Sol\`a 2009) which may have testable phenomenological
consequences for the current universe (Basilakos, Plionis \& Sol\`a 2009;
Grande et al. 2011).

The scenario proposed here can be viewed as a further step in the
direction of trying to find a cosmological framework based on the
fundamental principles of physics, and capable to link the dynamics of the
early universe with that of our late universe. The scenario is termed
complete in the sense that the model starts from a non-singular
inflationary stage which has a natural (universal) ending into the
radiation phase (thereby solving the horizon and  graceful exit problems),
and, finally,  the small current value of the vacuum energy density can be
conceived as a result of the massive disintegration of the vacuum into
matter during the primordial stages. In brief, our model evolves between
two extreme de Sitter phases. At early times the de Sitter phase is
unstable and drives the model continuously to a late time de Sitter state
driven by the remnant vacuum energy density.  As we shall see, the
leitmotiv of our approach is related to the structure of the effective
action of QFT in curved spacetime, which suggests that at early times only
even powers of the the Hubble parameter can contribute to a time varying
$\Lambda(t)$-term (see, for instance, Shapiro \& Sol\`a 2009).

The structure of the paper is as follows. In section 2, we briefly discuss
the background cosmological equations. In section 3, we discuss the basic
ideas underlying the time varying vacuum in an expanding Universe and set
up the basic equations whose solutions describe the complete evolution of
our model. In sections 4 and 5 we present the solutions in the early and
late universe respectively, while in section 6 we discuss our results and
at the same time we present some ideas towards extending the current
vacuum model. Finally, the main conclusions are summarized in section 7.

\section{Cosmology with a time dependent vacuum}

Let us recall that the cosmological constant contribution to the curvature
of space-time is represented by the $\Lambda\,g_{\mu\nu}$ term on the
\textit{l.h.s.} of Einstein's equations. The latter can be absorbed on the
\textit{r.h.s.} of these equations
\begin{equation}
R_{\mu \nu }-\frac{1}{2}g_{\mu \nu }R=8\pi G\ \tilde{T}_{\mu\nu}\,,
\label{EE}
\end{equation}
where the modified $\tilde{T}_{\mu\nu}$ is given by
$\tilde{T}_{\mu\nu}\equiv T_{\mu\nu}+g_{\mu\nu}\,\rL $. Here
$\rho_{\Lambda}=\Lambda/(8\pi G)$ is the vacuum energy density associated
to the presence of $\CC$ (with pressure $p_{\Lambda}=-\rho_{\Lambda}$),
and $T_{\mu\nu}$ is the ordinary energy-momentum tensor of 
matter and radiation. Modeling the expanding universe as a perfect fluid
with velocity $4$-vector field $U_{\mu}$, we have
$T_{\mu\nu}=-p_{m}\,g_{\mu\nu}+(\rho_{m}+p_{m})U_{\mu}U_{\nu}$, where
$\rho_{m}$ is the density of matter-radiation and
$p_{m}=\omega_{m} \rho_{m}$ is the corresponding pressure. Clearly the
modified $\tilde{T}_{\mu\nu}$ defined above takes the same form as
${T}_{\mu\nu}$ with $\rho_{\rm tot}=\rho_{m}+\rho_{\Lambda}$ and $p_{\rm
tot}=p_{m}+p_{\Lambda}=p_{m}-\rho_{\Lambda}$, that is
$\tilde{T}_{\mu\nu}=-p_{\rm tot}\,g_{\mu\nu}+(\rho_{\rm tot}+p_{\rm
tot})U_{\mu}U_{\nu}$ or explicitly,
\begin{equation}
\tilde{T}_{\mu\nu}= (\rho_{\Lambda}-p_{m})\,g_{\mu\nu}+(\rho_{m}+p_{m})U_{\mu}U_{\nu}\,.
\label{Tmunuideal}
\end{equation}
By assuming this generalized energy-momentum tensor and a spatially flat
FLRW metric,  the independent gravitational field equations reduce to
(Carvalho \& Lima 1992; Lima \& Maia 1994; Basilakos 2009; Sol\`a 2011) \be
 8\pi G\rho_{\rm tot}\equiv 8\pi G \rmr+\Lambda = 3H^2 \;,
\label{friedr} \ee

\be 8\pi G p_{\rm tot} \equiv   8\pi G \pmr-\Lambda =-2{\dot H}-3H^2\;,
\label{friedr2} \ee where the overdot denotes derivative with respect to
cosmic time $t$.

Now let us discuss a bit more the possibility that the vacuum energy
density is a function of the cosmic time. This is allowed by the
Cosmological Principle embodied in the FLRW metric. The Bianchi identities
(which insure the covariance of the theory) then imply
$\bigtriangledown^{\mu}\,{\tilde{T}}_{\mu\nu}=0$. With the help of the
FLRW metric, the previous identity amounts to the following generalized
local conservation law:
\be \dot{\rho}_{m}+\dot{\rho_{\Lambda}}+ 3H(\rho_{m}+p_{m}+
\rho_{\Lambda}+p_{\Lambda})=0\,. \label{frie2} \ee
Notice that we keep $G$ strictly constant, and therefore the assumption
$\dot{\rho_{\Lambda}}\neq 0$ necessarily requires some energy exchange
between matter and vacuum, e.g. through vacuum decay into matter, or vice
versa\,\footnote{There exists also the possibility that the vacuum is time
evolving and nevertheless non-interacting with matter. In this case,
however, either the DE has another component apart from $\CC$-- see the
$\CC$XCDM framework of (Grande, Sol\`a \& \v{S}tefan\v{c}i\'{c} 2006) --
or Newton's coupling is also time-varying, i.e. $\dot{G}\neq 0$ (Sol\`a
2008).}.


Let us remark that the EOS of the vacuum energy density maintains the
usual form $p_{\Lambda}(t)=-\rho_{\Lambda}(t)=-\Lambda(t)/8\pi G$ despite
the fact that $\Lambda(t)$ evolves with time. Inserting
$p_{\Lambda}=-\rho_{\Lambda}$ and $p_{m}=\omega_{m}\rho_{m}$ into
Eq.(\ref{frie2}) the latter equation leads to the following energy
exchanging balance between matter and vacuum:
\begin{equation}
\dot{\rho}_{m}+3(1+\omega_{m})H\rho_{m}=-\dot{\rho_{\Lambda}}\,. \label{frie33}
\end{equation}
Finally, combining equations (\ref{friedr}) and (\ref{frie33}), we find:
\begin{equation}
\dot{H}+\frac{3}{2}(1+\omega_{m}) H^{2}=4\pi G(1+\omega_{m})\rho_{\Lambda}=
\frac{(1+\omega_{m})\Lambda}{2}\,.
\label{frie34}
\end{equation}

\section{Running vacuum models evolving as a power series of $H$}
In what follows we investigate the cosmic expansion within a class of time
evolving vacuum models along these lines (Sol\`a 2011), and at the same
time involving ingredients capable of yielding a smooth transition from an
early de Sitter stage to a proper radiation and matter epochs (Lima \&
Maia 1994; Lima \& Trodden 1996). Consider the class of time evolving
vacuum models following an even power series of the Hubble rate:
\begin{equation}\label{powerH}
\CC(H)=c_0+c_2H^2+c_4H^4+c_6H^6...
\end{equation}
with $\rL(H)=\CC(H)/(8\pi\,G)$ the corresponding vacuum energy density.
The constant $c_0$ in (\ref{powerH}) represents the dominant term at low
energies (i.e. when $H$ is near the current value $H_0$). The $H^{2k}
(k\ge 1)$ powers represent small corrections to the dominant term and
provide a time evolving behavior to the vacuum energy density. The
expression (\ref{powerH}) has to be understood as a general ansatz for the
vacuum energy in an expanding universe. The reason for selecting even
powers of $H$ is because of the general covariance of the effective action
of QFT in a curved background. Although we cannot presently quote its
precise structure, we know it can only contain even powers which can
emerge from the contractions of the metric tensor with the derivatives of
the scale factor (Shapiro \& Sol\`a 2009). Interestingly, these
dynamical vacuum models have recently been linked with a potential
variation of the so-called fundamental constants of Nature (Fritzsch \&
Sol\`a 2012). The power like behavior with $H$ has also been seriously
considered as potentially emerging from a modification of Friedmann
equation from e.g. infinite-volume extra dimensions, and as an intriguing
and testable option for describing DE (Dvali \& Turner 2003).

The novelty in the present approach is that we extend the domain of
applicability of these models, namely we encompass here in a single
unified framework both the inflationary and the current dark energy
epochs. In a minimal model we expect that only a few powers of $H$ should
matter beyond the additive term, $c_0$, which is essential to insure a
good $\CC$CDM limit for any model of the sort (\ref{powerH}). While the
higher order powers $H^{2k}\, (k>1)$ are completely negligible at present,
they can acquire a great relevance in the early universe. This leads to
the following canonical realization of Eq.(\ref{powerH}) for describing
the basic features of both the early and late cosmos:
\begin{equation}
\label{RGlaw2} \Lambda(H)= c_0+3\nu H^2
+3\alpha\frac{H^4}{H_I^2}\,.
\end{equation}
Here $\nu$ and $\alpha$ are dimensionless parameters and the scale ${H_I}$
can be interpreted as the inflationary expansion rate (see below). With
only 2 additional parameters we need to choose between $\nu=0$ and the
existence of an early de Sitter phase connected with $\alpha\neq 0$. The
former choice means that at late times the model becomes indistinguishable
from $\CC$CDM at low redshifts. So, in order to have the initial de Sitter
phase and the possibility to confront our late time model with $\CC$CDM we
need at least 3 parameters. The third independent parameter has
conveniently been written as the ratio $\alpha/H_I^2$.

Inserting Eq.(\ref{RGlaw2}) into Eq.(\ref{frie34}) one can easily write
\begin{equation}
\label{HE}
\dot H+\frac{3}{2}(1+\wm)H^2\left[1-\nu-\frac{c_0}{3H^2}-\alpha\frac{H^2}{H_I^2}\right]=0\,.
\end{equation}
Remarkably there is the constant value solution  $H^2=(1-\nu)
H_I^2/\alpha$ of this equation for the very early universe (where we can
safely neglect $c_0/H^2\ll 1$). It signals the presence of an inflationary
epoch.

We shall present below the various phases of the decaying vacuum cosmology
(\ref{RGlaw2}), starting from an unstable inflationary phase [$a(t)
\propto e^{H_I t}$] powered by the huge value $H_I$ presumably connected
to the scale of a Grand Unified Theory (GUT) or even the Planck scale
$M_P$, then it deflates (with a massive production of relativistic
particles), and subsequently evolves into the standard radiation  and
matter dominated eras. Finally, it effectively appears today as a slowly
dynamical DE slightly correcting the standard $\Lambda$CDM model.

\section{From an early de Sitter stage to the standard radiation phase}
The Hubble function and scale factor of this model in the early universe
(when $c_0$ can be neglected) follows from direct integration of
Eq.(\ref{HE}):
\begin{equation}\label{HS1}
 H(a)=\left(\frac{1-\nu}{\alpha}\right)^{1/2}\,\frac{H_{I}}{\sqrt{D\,a^{3(1-\nu)(1+\wm)}+1}}\,,
\end{equation}
\begin{equation}\label{eq:afactor}
 \int_{a_{*}}^{a}\frac{d\tilde{a}}{\tilde{a}}\sqrt{D\,\tilde{a}^{3(1-\nu)(1+\wm)}+1}=\sqrt{\frac{1-\nu}{\alpha}}\,H_I\,\Delta t\,,
\end{equation}
where $D>0$ is a constant. Notice that $a_{*}$ is the scale factor at the
transition time ($t_{*}$) when the inflationary period ceases, and $\Delta
t=t-t_{*}$ is the cosmic time elapsed since then. Using Eq.(\ref{HS1}) and
the Einstein equations (\ref{friedr})-(\ref{friedr2}) we may also obtain
the corresponding energy densities:
\begin{eqnarray}\label{eq:densities}
&&{\rmr(a)}=\rho_I\,\frac{(1-\nu)^2}{\alpha}
\frac{Da^{3(1-\nu)(1+\wm)}}{\left[Da^{3(1-\nu)(1+\wm)}+1\right]^2}\\
&&\rL(a)=\frac{\CC(a)}{8\pi G}=\rho_I\,\frac{1-\nu}{\alpha}\
\frac{\nu Da^{3(1-\nu)(1+\wm)}+1}
{\left[Da^{3(1-\nu)(1+\wm)}+1\right]^2}\,,\nonumber\\
\end{eqnarray}
where $\rho_{I}=3H_{I}^{2}/8{\pi}G$ is the primeval critical energy
density associated to the initial de Sitter stage.

Obviously, if $Da^{3(1-\nu)(1+\wm)}\ll 1$ (i.e. $t\ll t_{*}$) then
Eq.(\ref{HS1}) boils down to the particular solution mentioned above, in
which $H=\sqrt{(1-\nu)/\alpha}\,H_I$ is constant, and $\rmr\simeq 0$,
$\rL\propto\rho_I$ (i.e. no matter and huge vacuum energy density). From
(\ref{eq:afactor}) it is obvious that $a(t)\sim
e^{\sqrt{(1-\nu)/\alpha}\,H_I\,\Delta t}$ and the universe then inflates.
For $Da^{3(1-\nu)(1+\wm)} \gg 1$ (i.e. $t\gg t_{*}$), instead,
Eqs.\,(\ref{eq:densities}) tell us that both $\rmr(a)$ and $\rL(a)$ decay
as $\sim a^{-3(1-\nu)(1+\wm)}$ while at the same time the ratio
$|\rL(a)/\rmr(a)|$ remains very small, viz. of order $|\nu|\leq{\cal
O}(10^{-3})$ (Basilakos, Plionis \&  Sol\`a 2009; Grande et al. 2011;
Basilakos, Polarski \& Sol\`a 2012), which insures that primordial
nucleosynthesis will not be harmed at all. Furthermore, since the vacuum
presumably decayed mostly into relativistic particles ($\wm=1/3$) we find
from Eq.\,(\ref{eq:afactor}) that in the post-inflationary regime $a\sim
t^{\frac{2}{3(1-\nu)(1+\wm)}}\sim t^{1/2}$, i.e. we reach essentially the
standard radiation epoch -- confirmed, in addition, by the fact that the
relativistic matter (and vacuum) energy
densities $\rmr(a)$ and $\rL(a)$ decay as $\sim
a^{-4}$ in this period. The universe thus evolves continuously from
inflation towards a standard (FLRW) radiation dominated stage, as shown in
the inner plot of Fig.\,1. In between these two eras, we see from the
first Eq.\,(\ref{eq:densities}) that we can have either huge relativistic
particle production $\rho_{r}=\rho_{m}\propto a^4$ in the
deflation period (namely
around $Da^{4} \ge 1$) or standard dilution
$\rho_{r}=\rho_{m} \propto a^{-4}$ well in the radiation era ($Da^{4} \gg 1$).

Naturally, due to the initial de Sitter phase, the model is free of
particle horizons. A light pulse beginning at $t=-\infty$ will have
traveled by the cosmic time $t$ a physical distance $d_{H}(t)=
a(t)\int_{-\infty}^{t}\frac{d\tilde{t}}{a(\tilde{t})}$, which diverges
thereby implying the absence of particle horizons: the local interactions
may homogenize the whole Universe. It should be clear that our accessible
part of the universe is only for times $t>t_{*}$, i.e. $\Delta t>0$. After
inflation has occurred and the FLRW (radiation dominated) phase has been
causally prepared, the cosmic time $\Delta t$ can just be called $t$ and
it is this one that parameterizes all cosmological equations in physical
cosmology (Peebles 1993).

\begin{figure}
\mbox{\epsfxsize=8.2cm \epsffile{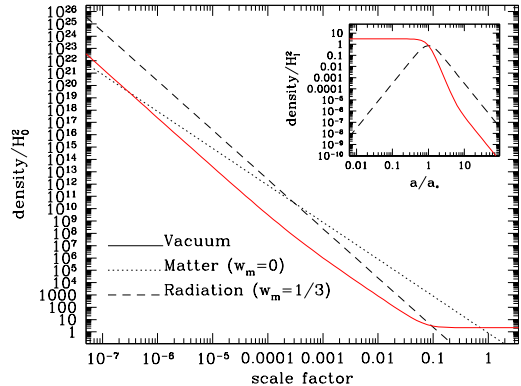}}
\caption{{\em Outer Plot:}
The evolution of the radiation, non-relativistic matter and vacuum
energy densities, for the unified vacuum model (\ref{RGlaw2})
in units of $H^{2}_{0}$. The curves shown are:
radiation (dashed line), non-relativistic matter (dotted line)
and vacuum (solid line, in red).
To produce the lines we used
$\nu=10^{-3}$, $\Omega^{0}_{m}=0.27$,
$\Omega^{0}_{R}=(1+0.227N_{v})\,\Omega^{0}_{\gamma}$,
$\Omega^{0}_{\Lambda}=1-\Omega^{0}_{m}-\Omega^{0}_{R}$,
$(N_{v},\Omega^{0}_{\gamma},h)\simeq
(3.04,2.47\times 10^{-5}h^{-2},0.71)$ (see Komatsu et al. 2011), and
set $\alpha=1$ and $D=1/a_{*}^{3(1-\nu)(1+\wm)}$.
{\em Inner Plot:}
the primeval vacuum epoch (inflationary period) into the
FLRW radiation epoch. Same notation for curves as before,
although the densities are now normalized with respect
to $H_I^2$ and the scale factor with respect to $a_{*}$ (see section 3).
For convenience we used $8\pi G=1$ units in the plots.}
\end{figure}


Although the motivation of the present model has a root in the general
structure of the effective action of QFT in curved space-time, we cannot
provide the latter at this point. However we can mimic it through a scalar
field ($\phi$) model for the interacting DE (Maia \& Lima 2002; Costa,
Alcaniz \& Maia 2008). This can be useful for the usual phenomenological
descriptions of the DE, and can be obtained from the usual correspondences
$\rho_{\rm tot} \rightarrow \rho_{\phi} =\dot{\phi}^{2}/{2} + V(\phi)$ and
$p_{\rm tot} \rightarrow p_{\phi} =\dot{\phi}^{2}/{2} - V(\phi)$
in Friedmann's Eqs.\,(\ref{friedr})-(\ref{friedr2}). We find ${4\pi
G}\dot{\phi}^{2}=-\dot{H}$ and
\begin{equation}
\label{Vz}
V=\frac{3H^{2}}{8\pi G}\left( 1+\frac{\dot{H}}{3H^{2}}\right)=\frac{3H^{2}}{8\pi G} \left( 1+\frac{1}{3}\frac{d\ln H}{d\ln a}\right) \,.
\end{equation}
We can readily work out the effective potential for our model
(\ref{RGlaw2}) in the early universe as a function of the scale factor.
Neglecting small ${\cal O}(\nu)$ corrections, which we have seen are not
important in the early stages, we arrive at
\begin{equation}
V(a)=\frac{\rho_I}{\alpha}\;\frac{1+Da^{4}/3}{(1+Da^{4})^{2}}.
\end{equation}
From this expression it becomes clear that the potential energy density
remains large and constant while $a\ll D^{-1/4}$ (i.e. before the
transition from inflation to the deflationary regime). Afterwards (when
$a\gg D^{-1/4}$) it decreases steadily as $V(a)\sim a^{-4}$, hence as
radiation. This confirms, in the effective scalar field language, the
previously described decay of the vacuum energy into relativistic matter
in our original framework.

\section{From the radiation-matter era to the residual dynamical dark energy at present}


Let us finally discuss the cosmic evolution after the inflationary period
is left well behind, i.e. when  $H\ll H_I$. The cosmic fluid will be first
in the radiation dominated epoch ($\wm=1/3$) and later on in the cold
matter dominated epoch ($\wm=0$). In this case (\ref{powerH}) reduces to
\begin{equation}\label{GeneralPS}
\CC(H)=\CC_0+3\,\nu\,(H^2-H_0^2)\,,
\end{equation}
where $\CC_0=c_0+3\nu\,H_0^2$ is the current value of the CC. Obviously
$c_0$ plays an essential role to determine its value, whereas the $H^2$
dependence gives some remnant dynamics even today, which we can use to fit
the parameter $\nu$ to observations. Using a joint likelihood analysis of
the recent supernovae type Ia data, the CMB shift parameter, and the
Baryonic Acoustic Oscillations one finds that the best fit parameters for
a flat universe are: $\Omega_{m}^0\simeq 0.27-0.28$ and $|\nu|={\cal
O}(10^{-3})$ (see Basilakos, Plionis \& Sol\`a 2009; Grande et al. 2011;
Basilakos, Polarski \& Sol\`a  2012). It is remarkable that the fitted
value of $\nu$ is within the theoretical expectations when this parameter
plays the role of $\beta$-function of the running CC (Shapiro \& Sol\'a
2002; Shapiro \& Sol\'a 2009; Sol\`a \& \v{S}tefan\v{c}i\'{c} 2005; Sol\`a
\& \v{S}tefan\v{c}i\'{c} 2006). In specific frameworks one typically finds
$\nu=10^{-5}-10^{-3}$ (Sol\`a 2008).

Despite the dynamical character of the vacuum energy (\ref{GeneralPS})
near our time, it is important to understand that a model of this kind
would not work for $c_0=0$, i.e. with only pure $H$-dependent terms. This
has been shown in Basilakos et al. (2009) and recently confirmed in Xu et
al. (2011), as the models of this sort fail to reproduce the observed CMB
and matter power spectrum. Furthermore, for them it is impossible to have
a transition from deceleration to acceleration (Basilakos et al. 2009). At
the root of all these problems is the fact that the $c_0=0$ models have no
$\CC$CDM limit.

The evolution equation for the Hubble function (\ref{HE}) in the
post-inflationary epoch reads:
\begin{equation}\label{eq:latetime}
 \dot H+\frac{3}{2}(1+\wm)(1-\nu)H^2-\frac{1+\wm}{2}\,c_0=0\,.
\end{equation}
Note that  $c_0$ cannot be neglected deep  in the cold matter dominated
period and specially near the present time. Trading the cosmic time for
the scale factor and using the redshift variable $1+z=1/a$ with the
boundary condition $H(z=0)=H_0$, one finds the solution of
Eq.\,(\ref{eq:latetime}) for the late stages ($\omega_m =0$)\footnote{If we use
three cosmic fluids namely non-relativistic matter, radiation and vacuum then
the Hubble function becomes
$$
H^{2}(z)= \frac{H_0^2}{1-\nu} \left[
\Omega_{m}^{0}(1+z)^{3(1-\nu)}+\Omega_{\Lambda}^0+
\Omega_{r}^{0}(1+z)^{4(1-\nu)}-\nu \right]
$$
where $\Omega_{m}^{0}=1-\Omega_{r}^{0}-\Omega_{\Lambda}^0$. Note that
$\Omega_{m}^{0}$ and
$\Omega_{r}^{0}$ are the non-relativistic and radiation density
parameters at the present time. Notice that using the cosmological data we
can put constraints on the free parameters
(see Basilakos et al. 2009; Grande et al. 2011).}:
\begin{equation}
\label{Hz}
 {H}^{2}(z)= \frac{H_0^2}{1-\nu} \left[(1-\Omega_{\Lambda}^0)(1+z)^{3(1-\nu)}+\Omega_{\Lambda}^0-\nu \right]\,,
\end{equation}
where
$\Omega_{\Lambda}^0=\Lambda_{0}/3H_0^{2}=8\pi\,G\,\rho_{\CC}^0/3H_0^{2}$.
In a similar way we can obtain the matter and vacuum energy densities as a
function of the redshift:
\begin{eqnarray}
\rmr(z)&=&\rmr^0{(1+z})^{3(1-\nu)}\,,\phantom{XXXXXXX} \\
\rho_{\Lambda}(z)&=&\rho^{0}_{\Lambda}+\frac{\nu\,\rmr^0}{1-\nu}\left[(1+z)^{3(1-\nu)}-1\right]\label{eq:rLz}.
\end{eqnarray}
From these equations it is clear that for $\nu=0$ we recover exactly the
$\CC$CDM expansion regime, the standard scaling law for nonrelativistic
matter and a strictly constant vacuum energy density $\rL=\rLo$ (hence
$\Lambda=\Lambda_0$). Recalling that $|\nu|$ is found to be rather small
when the model is confronted to the cosmological data, $|\nu|\leq{\cal
O}(10^{-3})$ (Basilakos et al. 2009; Grande et al. 2011), it is obvious
that at the present time this vacuum model is almost indistinguishable
from the concordance $\CC$CDM model, except for its mild dynamical
behavior which leads to an effective equation of state for the vacuum
energy that can mimic quintessence or phantom energy (Sol\`a 2011).  At
very late time we get an effective cosmological constant dominated era,
$H\approx H_0\,\sqrt{(\Omega_{\Lambda}-\nu)/(1-\nu)}$, that implies a pure
de Sitter phase of the scale factor.

\section{Discussion and future work}
Let us emphasize that in our model we need not specify the microscopic
nature of the matter fields involved, but in a more detailed formulation
baryons should obviously be conserved deep in the matter dominated epoch
and later in the vacuum phase. Thus we assume that the vacuum always
decays in the (material) dominant component at each phase. This means that
at late times the baryon-antibaryon decay is negligible, and, therefore,
we have an interaction in the dark sector, that is, the vacuum decays
mainly in cold dark matter (cf. Fritzsch \& Sol\`a 2012). In particular,
this means that the vacuum should not exceedingly decay into photons after
last scattering, as it could trigger an observable distortion of the CMB
spectrum. Fortunately, this does not happen if $\nu$ is as small as
$10^{-3}$. This has been studied in Opher \& Pelinson (2004; 2005) for
models of this sort, and is consistent with Basilakos et al. (2009)
results (see also Grande et al. 2011). It insures this framework does not
get in contradiction with basic facts of high precision cosmology.

In Figure 1 we display, in addition to the mentioned details of the early
stages of the cosmic evolution (inner plot), also the numerical analysis
describing the transition of the energy densities from the radiation epoch
into the matter dominated period, leading finally to the asymptotic de
Sitter phase beyond our time (outer plot).
The equality time of matter and radiation, $\rho_{m}(z_{\rm
eq})=\rho_{R}(z_{\rm eq})$ (Komatsu et al. 2011), is also marked in the
plot. It corresponds to $z_{\rm eq}\simeq 3200$, thus with no essential
change with respect to the $\CC$CDM.
On the other hand for $z\le 10$ (or $a\ge 0.1$) the vacuum energy density
appears as effectively frozen to its nominal value,
$\rho_{\Lambda}(z)\simeq \rho^{0}_{\Lambda}$, but still displaying a slow
cosmic evolution as in Eq.(\ref{eq:rLz}). The considered model therefore
provides a description of the cosmological vacuum as playing a prominent
role in the early (inflationary) universe and then evolving similarly to
the matter densities until it finally surfaces again at the present time.

We may wonder if the virtues of the cosmological picture under
consideration are confined to the peculiarities of the specific model
(\ref{RGlaw2}).  Remarkably, the latter is just the minimal or
``canonical'' implementation of the general class of models (\ref{powerH})
based on the even powers of $H$ which are favored from the point of view
of the general covariance. It turns out that many other models of this
class can still do the job. Interestingly, as it will be shown elsewhere,
all the vacuum models of the form (assuming $n$ integer):
\begin{equation}
\label{RGlawn} \Lambda(H)= c_0+3\nu H^2
+3\alpha\frac{H^{2n}}{H_I^{2n-2}}\ \ \ \ (n>1)
\end{equation}
(of which the model under study is just the particular case $n=2$ ) are
consistent with general covariance and perform automatically a successful
(``graceful exit'') transition from an unstable inflationary phase
(deflation) into the standard FLRW radiation dominated era, irrespective
of the differences in other details.

Another important aspect of our model is related to the ratio between the
early and  current values of  the vacuum energy density.
Understanding this ratio is the basic difficulty  defining the so-called
cosmological constant problem (Weinberg 1989).  For all values of the free
parameter $n$ appearing in the above equation, we see that value of
$\Lambda$ at the very early de Sitter phase is $\Lambda_I \sim  H^{2}_I$ while
at present it reads $\Lambda_0 \sim H^{2}_{0}$. Therefore, the ratio
between the two extreme vacuum energy densities reads
\begin{equation}
\frac{\rho_{vi}}{\rho_{vo}} \sim \frac{\Lambda_I}{\Lambda_0} \sim \frac{H^{2}_I}{H^{2}_{0}}\,.
\end{equation}
Notice that although $H_I$ (the  energy scale of the primeval de Sitter
state) is not given by the model, it can be estimated, for instance, from
the equilibrium temperature associated to the event horizon of an
arbitrary de Sitter state (Gibbons \& Hawking 1977). {In the present
case, it takes the following form: $k_B
T_I=\hbar(\Lambda_I/12\pi^2)^{1/2}$ ($k_B$ and $\hbar$ are the Boltzmann
and Planck constants). Therefore, using $H_I=\sqrt{\Lambda_I/3}$  we see
that the initial temperature of our scenario is $T_I={\hbar H_I}/({2\pi
k_B})$. Now, by choosing $H_I^{-1} $ to be of the order of the Planck
time, $t_P=\left(\hbar\,G/c^5\right)^{1/2}$, it is easy to check that the
initial temperature of the universe is $T_I=T_P/(2\pi)$, i.e. of order of
the Planck temperature, $T_P=\left(\hbar\,c^5/G\,k_B^2\right)^{1/2}$. More
important, for $H_I^{-1}$ of the order of the Planck time,  the model
essentially predicts that the above ratio is ${\rho_{vi}}/{\rho_{vo}} \sim
c^5/(\hbar\,G\,H_0^2)=\left(M_P\,c^2/H_0\,\hbar\right)^2$, where
$M_P=\left(\hbar c/G\right)^{1/2}$ is the Planck mass. In natural units
this simply reads ${\rho_{vi}}/{\rho_{vo}}\sim M_P^2/H_0^2\sim 10^{122}$,
where $M_P\sim 10^{19}$ GeV and $H_0\sim 10^{-42}$ GeV in such units.

The above result is consistent with the standard
theoretical expectation for the ratio between the
present day value of $\Lambda$  and its uppermost value at the
very early universe (Weinberg 1989).
In our model, such a result i.e. the present
smallness of the cosmological constant
($\Lambda_0 \sim 10^{-122} \Lambda_I$) can be seen
as the basic consequence of a decaying vacuum process in an aged Universe.
It is worth notice that the above ratio  was earlier
obtained for $n=1$ based on a similar approach, however,  not taking into
account that the action should be expressed in terms of even powers of the
effective action (Lima \& Maia 1994, Lima \& Trodden 1996). It should also be
stressed that the semi-classical value of $H_I$ provided by the
analogy with Gibbons-Hawking temperature relation holds
regardless of the value of $n$, and, in principle,  must be derived from a more fundamental QFT approach for a decaying vacuum
energy density. In addition,  one may also
think that the resulting concordance with the so-called $\Lambda$-problem
is not just a coincidence, and,  as such, it would be signalizing for a
deeper connection between the deflationary scenario and the quantum
gravity regime.

It should be also stressed that the higher order  $H^4$ term of our unified
ansatz (\ref{RGlaw2}), namely the term which is driving the early universe
dynamics, was motivated  from the covariance of the effective action. Let
us remark that our scenario is quite different from the
holographic-entropic models of inflation (Easson, Frampton \& Smoot 2012),
in which the thermal radiation from the temperature of the horizon is
associated to an energy density that varies as $\rho\sim T^4$, where $T$
is the aforementioned Gibbons \& Hawking temperature.  In particular, we
observe that at the very early times i.e., when $H=H_I$, there is no
radiation fluid component in our scenario. The universe starts its
evolution from a pure de Sitter state uniquely supported  by the vacuum
energy density.

On the other hand,  the existence of an early isothermal de Sitter phase
suggests that thermal fluctuations  may be the causal origin of the
primeval seeds that ultimately will form the galaxies. Such an origin
indicated by the present deflationary scenario are quite different from
the adiabatic fluctuations usually adopted in several variants of the
cosmic inflationary paradigm. In principle, one may expect that the
footprints of the primeval scale $H_I$ might  be extracted from CMB
observations trough the influence of the initial power spectrum of density
fluctuations on the  pattern of polarization and temperature anisotropies
at the last scattering surface. Such a possibility and its consequences
for the structure formation problem deserves a closer scrutiny and is
clearly out of the scope of the present paper.

It is also interesting that  the $H^2$ behavior at low energy,
complemented here by the contribution of the true vacuum constant term
$c_0$,  permits the smooth  transition from a deceleration to an
acceleration phase as required from Supernovae observations either in the
dynamics as well as in the kinematic descriptions (Turner \& Riess 2002;
Cunha \& Lima 2008; Guimar\~aes, Cunha  \& Lima 2009, Guimar\~aes \& Lima
2011).  As explained in detail by Basilakos, Polarski \& Sol\`a
(2012), the necessary transition is not possible in the decaying vacuum
models driven only by the dynamical $H^{2}$ term (or its variant
$\dot{H}$) without introducing the additive constant $c_0$, such as e.g.
in the entropic-force model of the current accelerated expansion (Easson,
Frampton \& Smoot 2011). Quite in contrast, for our unified model as
given by (\ref{RGlaw2}), as well as for the entire class of models
generated by the ansatz (\ref{RGlawn}), we obtain a consistent description
both at high energies, i.e. for the dynamics of the early inflationary
universe, and at low energies, i.e. for the physics of the present
universe. In this sense, our model can be considered as an effective
unified framework for a  practical description of the complete
history of our cosmos. Somehow it is the kind of framework that is
awaiting for a more fundamental theory whose effective behavior is of the
sort described here, since then all of the discussed cosmological problems
could be accounted for in an efficient way. In view of the fact that
our unified ansatz (\ref{RGlaw2}) has a structure which is consistent with
the general covariance of the effective action both at low and high
energies, we expect that this feature should pave the way for a smooth
connection of our framework with that fundamental theory.

\section{Conclusions}

In this work we have put forward a new class of global cosmological
scenarios which provides a consistent and rather complete account of the
expansion history of the Universe. Although only  partially justified from
a more fundamental approach, it can be conceived as providing the simplest
effective dynamical structure that the vacuum energy should inherit from
some future fundamental theory in order to automatically solve these
problems in an economical way. As we have emphasized throughout the
paper, the present dynamical vacuum framework based on the $H^2$ and
$H^4$ terms is compatible with the general form of the effective action of
QFT in curved space-time.

It was also pointed out the existence  of a large class of generalized
models of this sort (see Eq.\,\ref{RGlawn}), all of them sharing a similar
phenomenological behavior, namely: (i) the Universe starts from an
inflationary non-singular state, thus overcoming the horizon problem; (ii)
the early inflationary regime has a natural (universal) ending into the
radiation phase; and (iii) the small current value of the vacuum energy
density can be conceived as a result of the massive disintegration of the
vacuum into matter in the primordial stages. The upshot is a new unified
vacuum picture of the cosmic evolution spanning from the early inflation
period to the late dark energy era and deviating only very mildly from the
observed $\CC$CDM behavior. The amplitude of the expected deviations
($|\nu|$ $\sim 10^{-3}$) can decide which is the more realistic
description of the vacuum at late stages, and, potentially, it may provide
an indication favoring  the complete decaying vacuum scenario proposed
here. The extended scenario will be discussed in detail in a forthcoming
communication.

\section*{Acknowledgments}
JASL is partially supported by CNPq and FAPESP (Brazilian Research
Agencies). JS has been supported in part by MEC, FEDER, CPAN and by
2009SGR502 Gen. de Catalunya. SB thanks the Dept.\ ECM of the Univ.\ de
Barcelona for the hospitality when part of this work was being done.

\end{document}